
\input harvmac

\def\part{\partial}
\def\ssc{\scriptscriptstyle}
\def\DD{\hbox{D}}
\def\dd{\hbox{d}}
\def\ee{\hbox{e}}

\lref\zwie{T. Kugo, B. Zwiebach, ``Target space duality as a symmetry of
string field theory", IASSNS-HEP-92/3  (Jan. 1992)}
\lref\p{ T. Bhattacharya, A. Gocksch, C.P. Korthals Altes
and R.D. Pisarski, {\sl Phys. Rev. Lett.} {\bf 66} (1991) 998}
\lref\shs{ S.H. Shenker, in {\sl Random surfaces and quantum gravity},
ed. O. Alvarez et al., (New York, Plenum, 1990)}
\lref\lipa{ L.N. Lipatov, {\sl  Pis'ma Zh. Eksp. Teor. Fiz.} {\bf 25}
(1977) 116}
\lref\fs{ W. Fischler and L. Susskind, {\sl Phys. Lett.} {\bf B171}
(1986) 383; {\sl Phys. Lett.} {\bf B173} (1986) 262}
\lref\ds{Dine and Seiberg}
\lref\gregm{G. Moore, R. Plesser, S. Ramgoolam, {\sl Nucl. Phys.}
{\bf B377}(1992) 143}
\lref\ds{M. Dine, N. Seiberg, {\sl Phys. Lett.} {\bf 162B} (1985) 299}
\lref\zinn{J. C. LeGuillou, J. Zinn-Justin, {\sl Large
Order Behavior of Perturbation Theory}
(Amsterdam, North Holland, 1990)}
\lref\rob {R. Myers, {\sl Phys. Lett.} {\bf B199} (1987) 371}
\lref\ramy{ R. Brustein, B. Ovrut,
``Nonperturbative Interactions in
String Theory", UPR-524T, hep-th/9209033
(Aug. 1992);
``Nonperturbative Effects
in 2-D String Theory or Beyond the Liouville Wall", UPR-523T,
hep-th/9209081 (Sept. 1992);
``Stringy Instantons", UPR-523T, hep-th/9209045
(Sept. 1992)}
\lref\wadia{A. Dhar, G. Mandal, S. Wadia,
``A Time Dependent Classical Solution of c=1 String Field Theory
and Nonperturbative Effects", TIFR-TH-92-40A,
hep-th@xxx/9212027 (Dec. 1992)}
\lref\jev{A. Jevicki, {\sl Nucl. Phys.}{ \bf B376} (1992) 75}
\lref\modinv{M. Cvetic, F. Quevedo, S-J Rey, {\sl Phys. Rev. Lett.}
{\bf 67} (1991) 1836; M. Cvetic, S. Griffies, S-J Rey, {\sl Nucl. Phys.}
{\bf B389} (1993) 3}
\lref\jaje{J. Avan, A. Jevicki, {\sl  Phys. Lett.}
{\bf B266} (1991) 35}
\lref\wein{ For a recent review of related topics see E. Weinberg,
``Vacuum Decay in Theories with Symmetry
Breaking by Radiative Corrections",
CU-TP-577, hep-ph/9211314 (Nov. 1992)}
\lref\sen{A. Sen, ``Electric Magnetic Duality in String Theory",
TIFR-TH-92-41, IC-92-171, hep-th/9207053  (July 1992)}
\lref\sendy{A. Sen, ``Quantization of Dyon Charge and
Electric-Magnetic
Duality in String Theory", TIFR/TH/92-46,
hep-th@xxx/9209016 (Sept. 1992)}
\lref\sentor{ A. Sen, ``SL(2,R) Duality and
Magnetically Charged Strings",
TIFR-TH-93-03, hep-th/9302038 (Feb. 1993)}
\lref\sench{ A. Sen, ``Magnetic Monopoles,
Bogomol'nyi bound and SL(2,Z)
Invariance in String Theory", NSF-ITP-93-29,
TIFR-TH-93-07, hep-th/9303057
(March 1993)}
\lref\anam{ A. Font, L. E. Ibanez, D. Lust,
F. Quevedo, {\sl Phys. Lett.}
 {\bf B249} (1990) 35}
\lref\duff{ M. Duff, {\sl Class. Quantum Grav.} {\bf 5} (1988) 189}
\lref\dav{F. David, {\sl Nucl. Phys.} {\bf B348} (1991) 507}
\lref\strom {A. Strominger, {\sl Nucl. Phys.} {\bf B343} (1990) 167 }
\lref\otherfive{
M. J. Duff and J. X. Lu, {\sl Nucl. Phys. }{\bf B354} (1991) 129,
{\bf B354} (1991) 141,
{\sl Phys. Rev. Lett.} {\bf 66} (1991) 1402;
C. Callan, J. Harvey, A. Strominger, {\sl Nucl. Phys.}
{\bf B367} (1991) 60}
\lref\schw {J. Maharana and J.H. Schwarz, {\sl Nucl. Phys.}
{\bf B390} (1993) 3; J. Schwarz, ``Dilaton-Axion Symmetry'',
CALT-68-1815, hep-th@xxx/9209125
(Sept. 1992) }
\lref\dyon{  A. Shapere, S. Trivedi, F. Wilczek, {\sl Mod. Phys. Lett.}
{\bf A6} (1991) 2677}
\lref\mm{
E. Brezin, V. Kazakov, {\sl Phys. Lett.} {\bf B236} (1990) 144;
M. Douglas, S.H. Shenker, {\sl Nucl. Phys.} {\bf B335} (1990) 635;
D. Gross, A. Migdal, {\sl Phys. Rev. Lett.}
{\bf 64} (1990) 127}
\lref\djp{S. Das, A. Jevicki, {\sl Mod. Phys. Lett. }
{\bf A5} (1990) 1639;
J. Polchinski,{ \sl Nucl. Phys.} {\bf B346} (1990) 253}
\lref\paulm{J. Lee and P. Mende, ``Semiclassical Tunneling in
(1+1) Dimensional
String Theory", Brown-HET-880, hep-th/9211049 (Oct 1992)}
\def\de{\delta}
\Title{\vbox{\baselineskip12pt{\hbox{iassns-hep-93-24}}%
{\hbox{fermilab 93/087-T}}{\hbox{nsf-itp-93-45}}{\hbox{hep-th/9304129}}%
{\hbox{revised}}}}
{\vbox{\centerline{The Strength of String Nonperturbative Effects}
\vskip2pt\centerline{and}
\vskip2pt\centerline{Strong-Weak Coupling Duality}}}
\bigskip
\centerline{J.D. Cohn${}^\dagger$\footnote{*}{jdcohn@fnalv.fnal.gov;
Address after Sept.1, 1993: Dept. of Physics, Univ. of California, Berkeley,
CA, 94720}}
\centerline{M.S. 106, Fermilab, P.O. Box 500, Batavia, Illinois 60510}
\centerline{and Center for Theoretical Physics, MIT, Cambridge, MA  02139}
\medskip
\centerline{and}
\medskip
\centerline{Vipul
Periwal${}^\dagger$\footnote{**}{vipul@guinness.ias.edu}}
\centerline{The Institute for Advanced Study,
Princeton, New Jersey 08540-4920}
\bigskip
\centerline{${}^\dagger$Institute for Theoretical Physics, UCSB}
\centerline{Santa Barbara, California  93106-4030}
\bigskip
\centerline{ABSTRACT}
A strong-weak coupling duality symmetry
of the string equations of motion has been suggested in the
literature.
This symmetry implies that vacua occur in pairs.
Since the coupling constant is a dynamical variable in string theory,
tunneling solutions between strong and weak coupling vacua may exist.
Such solutions would naturally lead to nonperturbative effects with
anomalous coupling dependence.
A highly simplified example is given.
\bigskip
\vfill\eject
Matrix model descriptions\mm\ of strings
in low dimensions provide
tractable generating functionals of string perturbation theory.
At large orders, the perturbation theory for these strings grows
at order $k$
as $(2k)!$ rather than the usual $k!$ found in most field theories.
Shenker\shs\ has presented arguments based on properties of
the moduli spaces of Riemann surfaces with $k$ handles that this
growth is generic.
However, a spacetime interpretation of the origins of the
nonperturbative effect underlying this growth is not known.
This is the issue we address here.

In field theory,
the growth of perturbation theory is related to the coupling constant
dependence of nonperturbative effects\zinn.  Starting with
$Z(g)\equiv\int \DD\psi \ee^{-S}\equiv \sum g^k Z_k ,$
Lipatov's argument\lipa\ applies
saddle point analysis to
${Z_k = \oint \dd g g^{-k-1} Z(g)}$ for large $k$.
For
${S[\psi,g] = \int \dd^Dx \left[(\part\psi)^2 + g^{n-1}\psi^{2n}\right], }$
the saddle point equations for $g_s, \psi_s$ are:
\eqn\saddles{\eqalign{{k+1\over g_s}  +
(n-1)g_s^{n-2}\int \psi_s^{2n}&=0,\cr
-\part^2\psi_s +ng_s^{n-1}\psi_s^{2n-1} = 0.\cr }}
Rescaling $\sqrt{-g} \psi = \phi,$
one finds $Z_k \sim g_s^{-(k+1)} e^{-S(\phi_s,1)/g_s},$ where
$S(\phi_s,1)$ is independent of $k$,
\eqn\gsad{g_s = {1\over k+1}\int \phi_s\part^2\phi_s {n-1\over n},}
and the l.h.s. of this equation is
negative definite if $\phi_s$ is real.
Eq. \gsad\ implies that $Z_k$ grows as
$k!,$ up to factors of the form $k^\alpha A^k$ which come from taking
into account fluctuations about, and
zero modes of, the saddle point configuration.
(Of course, $Z_k$ is only nonzero in this example when
$k=0 \; \hbox{mod} \, (n-1).$)
For a more general potential the argument goes through similarly.
The above explicitly shows that $\phi_s,$ the saddlepoint solution,
is independent of $k,$ so the growth in the perturbation series is due
to the explicit $1/g$ in front of the rescaled action.

In the case of an effective action, write
$\psi \equiv \tilde{\psi} + \xi$, and integrate out
the fluctuations $\xi$ above some mass scale to get
$Z(g) = \int \DD
\tilde{\psi} \ee^{-S_{{\ssc \rm eff}}(\tilde{\psi} )}$.
Its saddle point is at $\tilde{\psi}$ such that
$\de S_{{\ssc\rm eff}} /\de \tilde{\psi} = 0$.
Schematically then, ignoring zero modes,
\eqn\effg{{k+1\over g_s}  =- {\part S_{{\ssc\rm eff}}\over
{\part g}}|_{g_s,\tilde{\psi}_{s}},}
may be expected to reproduce the same $k!$ growth as discussed
above.  However, quantum effects
(due to the fluctuations that have been
integrated out) may
induce a change in the vacuum structure\wein\ and thus allow for new
solutions with anomalous $g$ dependence.
A recent example, due to  Bhattacharya {{\it et al.}}\p,
shows in the context of an $SU(N)$ gauge theory at high
temperature that such a change in vacuum structure due to radiative
corrections does indeed
lead to anomalous $g$ dependence of the interface tension of domain
walls between quantum vacua.
Calling the string coupling constant $\kappa$,
the $(2k)!$ growth in string theory would follow from an
instanton action scaling as $\kappa$.  One sees
from the above that
explicit terms of $O(\kappa)$ in the
action may not be necessary to produce this.

We suggest that a strong-weak coupling
duality in string theory\foot{This
has recently been discussed in several string backgrounds
\strom\anam\otherfive\dyon\sen\schw. It
generalizes a duality found in supergravity,
for extensive references see \sen.}
may lead to nonperturbative effects with anomalous $\kappa$ dependence,
as follows.  This duality interchanges strong and weak coupling, and the
r\^oles of particles and solitons.  It leaves the equations of motion
invariant, but not the action.  Thus solutions to the equations of motion,
{\it e.g.}, vacua of string theory, are related by this symmetry and
generically appear in pairs.  Because
the duality involves the loop counting
parameter, the multiple vacua will only
become evident in the equations of
motion when quantum corrections are included.
The coupling constant in string theory is a dynamical variable,
so there can be field configurations that
interpolate between such pairs of vacua.
The actions of such configurations will scale anomalously with what
we interpret in (one of) the weak coupling
vacuum(vacua) as the coupling.

The concepts of tunneling and anomalous scaling must be made
precise for these configurations, since they involve changing the
dilaton field whose expectation value determines the coupling.
In this form of duality, physics in the strong coupling vacuum
has a description in terms of weakly
coupled solitons. A complete tunneling
process should be described as tunneling from the weak coupling vacuum
($\kappa$ small)
until a point in field space where $\kappa=O(1),$ after which
the appropriate description is one of tunneling into the strong coupling
vacuum ($\kappa^{-1}$ small), described in terms of the weakly coupled
dual theory.  In a given weak coupling vacuum, the effective
action is a series in some small parameter.  Nonperturbative effects
usually have actions $\sim O(1/\kappa^2)$.
Here, in a given perturbative
vacuum one can identify the coupling.  What is meant by anomalous
scaling is that the actions of the instantons will not
be $\sim O(1/\kappa^{2}).$

This can be demonstrated in a toy model, putting the system
in finite volume and neglecting gravity.  The scaling
is anomalous and depends upon the details of the potential.
The duality symmetry appears in the equations of motion
for the action in the
Einstein basis, where the metric has been
rescaled so that the coupling does not appear out in front of the
action.
Leaving out gravity and
other fields, the action is
\eqn\simpa{
S = \int \dd^{D-1} x \dd t
\left[ ({{\part \kappa}\over \kappa})^2 - V(\kappa) \right]
}
In finite spatial volume of size $L^{D-1}$
a time dependent classical solution to the equations of
motion has finite action\foot{JDC
thanks J. Polchinski for discussions about this.}.
Searching for a solution $\kappa = \kappa(t)$ and using energy conservation,
\eqn\econ{
\left({1 \over \kappa}{{\dd\kappa}\over {\dd t}}\right)^2
+ V(\kappa) = E \; \; ,
}
choose $V(\kappa) = g(\kappa)^2 , g(\kappa) = g(1/\kappa)$.  The extrema
of $V$ are at $2 g(\kappa)g'(\kappa) = 0$.  Consider a path
starting at $\kappa_0$ and ending at $1/\kappa_0$, where
$g, \kappa_0$
obey
$g(\kappa_0) = g(1/\kappa_0) = V(\kappa_0) = E = 0$.  Then from conservation
of energy, for this classical solution, ${{\dd\kappa}/{\dd t}} =
 \kappa \sqrt{-V(\kappa)}$ and
\eqn\sclass{
S_{cl} = - \int \dd^{D-1}x dt \ 2 V(\kappa)
= - L^{D-1} \int {\dd \kappa}\ {{\dd t }\over \dd\kappa}\ 2 V(\kappa)
=  2i L^{D-1} \int_{\kappa_0}^{1/\kappa_0} {\dd\kappa \over \kappa} g(\kappa)
}
For a potential $g(\kappa) = \sum a_n (\kappa^n + \kappa^{-n})$,
\eqn\sfin{
S_{cl} =  4 L^3 i \left[ \sum_{n > 0} {a_n \over n}(\kappa_0^{-n }-
\kappa_0^{n}) - 2 a_0 \ln \kappa_0 \right]
}
The simplest potential with an extremum away from $\kappa = \pm 1$ is
$V = (\kappa + 1/\kappa - a)^2$.  The minima (couplings in the
perturbative vacua) are
$ \kappa_{\pm} = (a/2)\left[1 \pm \sqrt{1 - 4/a^2}\right]. $
The corresponding action, when $a$ is large,
scales to leading order as $1/\kappa_-$,
that is, in theories where the coupling $\kappa_-$ is
small.  The coupling constant dependence of the action evaluated
at this solution receives corrections
from the measure due to zero modes, altering {\it e.g.} the logarithmic
dependence on the coupling.
Solutions to the equations of motion which keep the dilaton fixed
have actions that scale as $\kappa^{-2}$ as usual.

If strong-weak coupling duality is
generic to string theory, an understanding of
it should be based on a stringy formulation (such as can be done for
$R \leftrightarrow 1/R$ duality, and its generalizations, in string field
theory\zwie).  So far,
supporting arguments for this duality have been
presented that are specific to
particular string backgrounds\strom\anam\sen\schw.
A coupling constant duality has been conjectured\strom\
in ten dimensional string--five-brane duality\duff\strom.
In the case of compactification of the heterotic string on a
six dimensional torus, it was found that
dyonic solutions to the equations of
motion were duality invariant\dyon, and
Sen\sentor\ showed that this duality is a full symmetry of the equations
of motion (of the low energy weak coupling lagrangian).  He
also showed that the spectrum of the charges of the particles and
the solitons is
consistent with this duality\sendy\sench.

Jevicki\jev\ has also found a duality
between the solitons and particles in the
$c=1$ collective field theory of matrix models with no matrix model
potential\foot{This theory maps\jaje\ to the
$c=1$ collective field theory with a matrix
model potential and a time dependent chemical potential.}.
It is a symmetry of the equations
of motion but not of the action.  It is not clear whether
this is a strong-weak coupling duality
because the spacetime interpretation
is unknown.

A calculable example is required
to see the consequences of nonperturbative
effects due to the pairing of vacua
caused by strong-weak coupling duality.
It is necessary to have a tractable
description of the theory for the regime where
$\kappa \sim O(1)$.
One difficulty with computations in
the known examples is the requirement that one explicitly
express  the fields in the dual vacua in terms of
each other, $\Phi_{dual} = \Phi_{dual}[\Phi].$   This must be done,
when $\kappa \sim O(1),$
in order to match tunneling solutions coming out of/going in to
the dual vacua.
The examples with $N=4$ supersymmetry have
strongly constrained radiative corrections
and may allow calculations of tunneling effects.
Sen\sench\ points out that the tree-level spectrum of the strongly
coupled theory
may be preserved in some cases
due to supersymmetry nonrenormalization theorems.
Ref.~\anam\ pointed out several analogies
between $R\leftrightarrow 1/R$ duality and
strong-weak coupling duality.  For $N=1$ supergravity in
four dimensions they
wrote down the most general superpotential for the field
$S = e^{\phi} + i a$ ($\phi$ is the dilaton, $a$ is the axion)
which is
without singularities for finite values of $S$.
It may be possible to
have more general duality invariant couplings for the kinetic terms
and couplings to other fields.
Neglecting gravity and other fields besides
the axion, and putting the system in a finite volume, these
potentials can be treated as above (or by using \modinv).
For time dependent solutions
interpolating between a supersymmetry preserving
minimum at $S_{min}$ and its dual, the
dependence upon $S_{min}$ goes as
$|e^{i \theta} W(S_{min})(1 - e^{i \beta}S_{min})|$ times the volume.
Here $\beta, \theta$ are phases, and $\theta$ is chosen to maximize the
action.   Ref.~\anam\ points out that duality symmetry
may prevent an arbitrarily weak coupling expansion from being valid,
see also Ref.~\ds.

It is not clear how to connect
strong-weak coupling duality to the source
of nonperturbative effects which has been identified\shs\ in solvable
matrix models of string theory.  The bare coupling constant
is $N^{-2}.$ The models reduce to the study of the interactions of
$N$ eigenvalues.  The leading nonperturbative effects are tunnelings of
individual eigenvalues\dav\shs\ in the matrix model potential.
Since $1/N$ of the degrees of freedom
are involved, the motion results in
an instanton action $\sim e^{-S_{inst}N^2} \sim e^{- {\rm (const.)}N}.$
There have been some attempts to describe these in the
string field theory for the $c=1$
matrix model\foot{Ref.~\ramy\ describes
the $c=1$ matrix model tunneling in the collective field theory\djp,
emphasizing that the coupling must appear explicitly
in the action.  Ref.~\paulm\ uses a minimal phase space
volume argument to constrain the instanton size.  Ref.~\wadia\
shows that a quantum wavepacket
has some support in the classically forbidden
region.}.   There are many
nonperturbative definitions of the $c=1$ theory\gregm.
One possibility, if there exist boundary conditions which
respect this duality symmetry, is to
require\foot{We thank M. Douglas for suggesting this.}
duality symmetry to be respected by the nonperturbative
definition of the model.
\medskip
Acknowledgements: We thank A. Sen, S. Shenker and A. Strominger
for discussions. J.D.C. also thanks
O. Alvarez, A. Anderson, T. Banks,
L. Brekke, S. Cordes, M. Douglas, D.E. Freed, D. Kastor, J. Lykken,
J. Polchinski, P. Townsend and J. Traschen.
V.P. thanks I. Klebanov and R. Myers.
V.P. was supported by grant no. DE-FG02-90ER40452.
This research was
supported in part by the National Science Foundation under Grant No.
PHY89-04035.  J.D.C. thanks
R. Brooks for hospitality at MIT.
\bigskip

\listrefs

\bye